\documentclass[
    final         % use final for the camera ready runs
 ]
  {aipproc}
\layoutstyle{8x11single}
\setcitestyle{numbers,square}
\begin{document}

\title{High-mass twins \& resolution of the reconfinement, masquerade and hyperon puzzles of compact star interiors}


\author{David Blaschke}{
  address={Bogoliubov Laboratory for Theoretical Physics, JINR Dubna, 141980 Dubna, Russia}
  ,altaddress={Instytut Fizyki Teoretycznej, Uniwersytet Wroclawski, 50-204 Wroclaw, Poland} % additional visiting address
}

\author{David E. Alvarez-Castillo}{
  address={Bogoliubov Laboratory for Theoretical Physics, JINR Dubna, 141980 Dubna, Russia}
  ,altaddress={Instituto de F\'{i}sica, Universidad Aut\'{o}noma de San Luis Potos\'{i}, San Luis Potos\'{i}, 
  M\'{e}xico} % additional visiting address
}

\begin{abstract}
We aim at contributing to the resolution of three of the fundamental puzzles related to the still unsolved problem of the structure of the dense core of compact stars (CS): (i) the hyperon puzzle: how to reconcile pulsar masses of $2\,$M$_\odot$ with the hyperon softening of the equation of state (EoS); (ii) the masquerade problem: modern EoS for cold, high density hadronic and quark matter are almost identical; and (iii) the reconfinement puzzle: what to do when after a deconfinement transition the hadronic EoS becomes favorable again? We show that taking into account the compositeness of baryons (by excluded volume and/or quark Pauli blocking) on the hadronic side and confining and stiffening effects on the quark matter side results in an early phase transition to quark matter with sufficient stiffening at high densities which removes all three present-day puzzles of CS interiors.
Moreover, in this new class of EoS for hybrid CS falls the interesting case of a strong first order phase transition which results in the observable high mass twin star phenomenon, an astrophysical observation of a critical endpoint in the QCD phase diagram.
\end{abstract}

\maketitle

%%%%%%%%%%%%%%%%%%%%%%%%%%%%%%%%%%%%%%%%%%%%
%% MAINMATTER
%%%%%%%%%%%%%%%%%%%%%%%%%%%%%%%%%%%%%%%%%%%%

\section{Introduction}

Neutron stars (NS)  are unique  cosmic laboratories \cite{Weber:1999qn,Glendenning:2000} to study matter and space-time under the most extreme  conditions not attainable in terrestrial experiments.  
Explaining observations  of  NS  is a challenge to theorists: the density inside NS can be an order of magnitude larger than that in atomic nuclei.  
On the other hand,  NS  give a unique chance to test our theories of the fundamental structure of matter  
via confronting models  based on these theories with NS observations. 
A challenge for the description of dense matter   in neutron stars (NS) and supernovae (SN) is the appearance of exotic particle species, namely hyperons and deconfined quarks \cite{Alford:2006vz}. 
Presently, it is not precisely known which conditions in terms of temperature, density and iso-spin asymmetry serve as a threshold for their appearance or disappearance. 
A particularly interesting situation arises from the fact that model calculations imply that these thresholds 
for the occurrence hyperons and deconfined quark matter are similar if not overlapping, thus being a source of ambiguities concerning the equation of state (EoS)  of dense matter. 
In what follows we shall discuss just three of many possible types of neutron star interiors:
pure NS, NS with hyperonic cores and NS with quark matter cores. All these superdense objects 
belonging to the class of compact stars (CS).
 
\begin{figure}
  \includegraphics[height=.35\textheight]{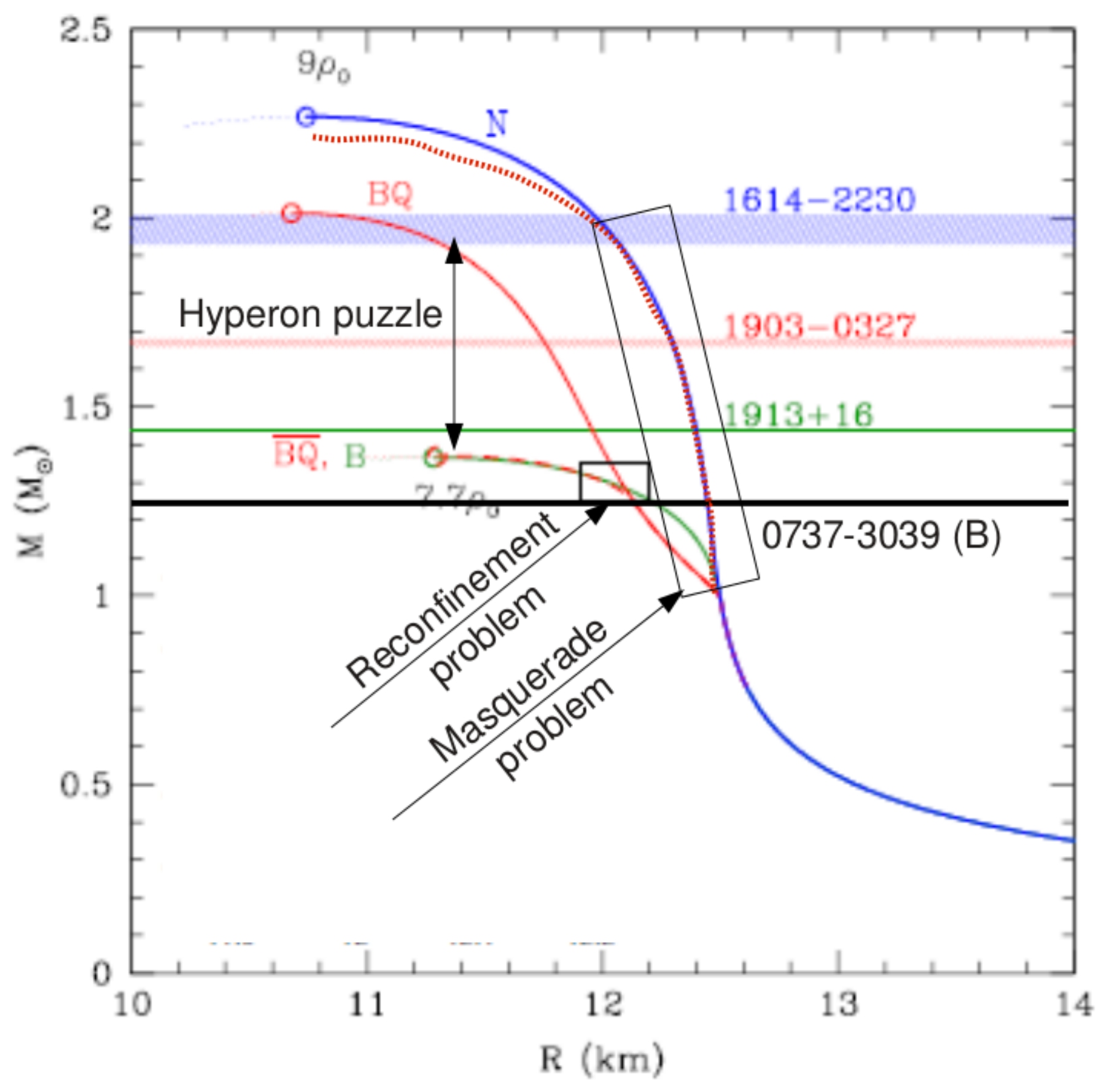}
  \includegraphics[height=.35\textheight]{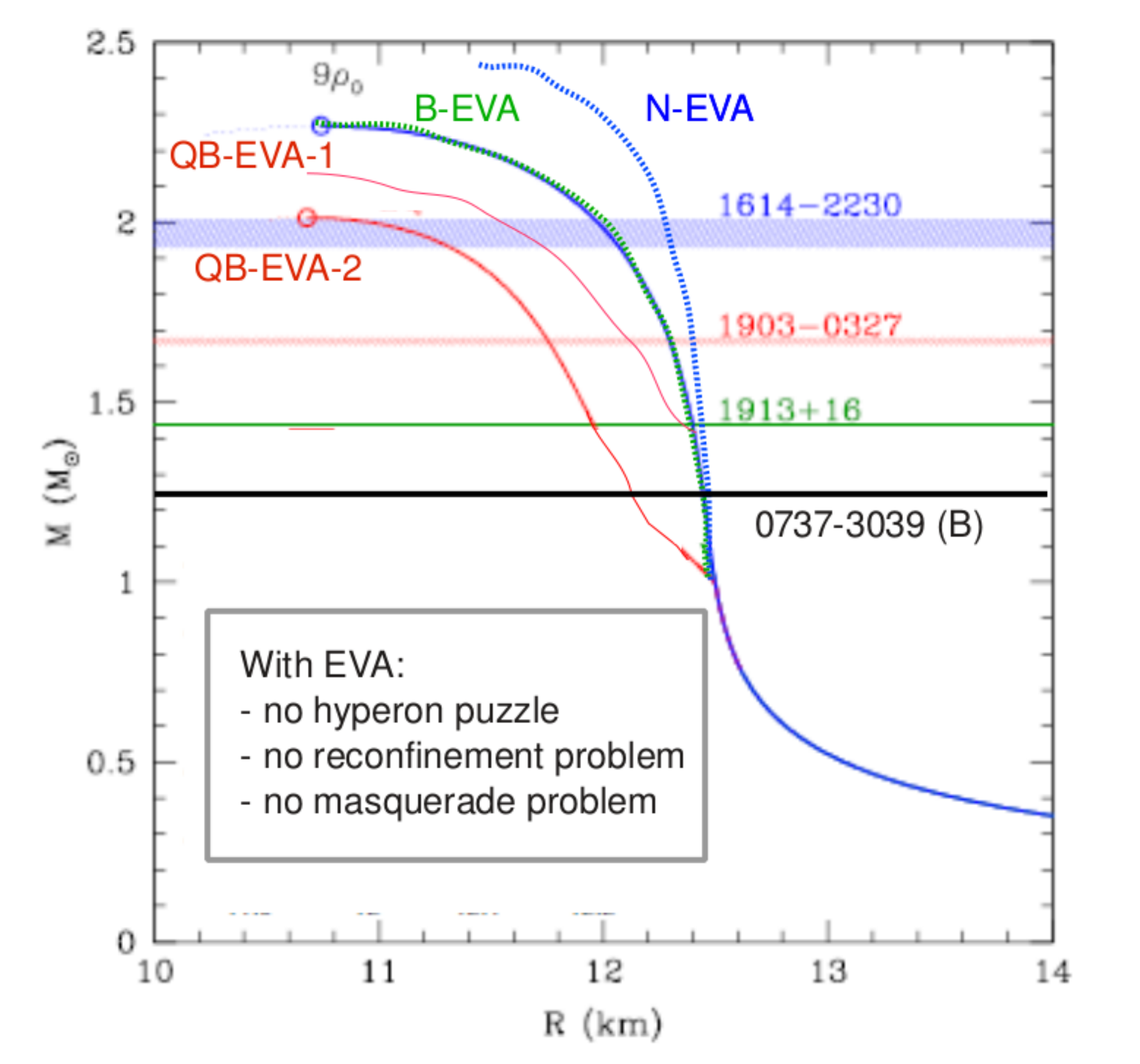}
  \caption{Mass-radius sequences for different model equations of state (EoS) illustrate how the three major problems in the theory of exotic matter in compact stars (left panel) can be solved (right panel) by taking into account the baryon size effect within a excluded volume approximation (EVA). Due to the EVA both, the nucleonic (N-EVA) and hyperonic (B-EVA) EoS get sufficiently stiffened to describe high-mass pulsars so that the hyperon puzzle gets solved which implies a removal of the reconfinement problem. Since the EVA does not apply to the quark matter EoS it shall be always sufficiently different from the hadronic one so that the masquerade problem is solved. 
  \label{fig:1}}
\end{figure}

At present any attempt to unravel the true composition of the interior of CS has to be confronted with
the observation of two CS with a mass of $2~$M$_\odot$ \cite{Demorest:2010bx,Antoniadis:2013pzd}. 
Hereby the following three puzzles arise:
\begin{itemize}
\item \underline{the hyperon puzzle:} 
how to reconcile pulsar masses of $2\,$M$_\odot$ with the hyperon softening of the equation of state (EoS)
which limits the maximum mass of NS with hyperons to much lower values \cite{Baldo:2003vx}; 
\item \underline{the masquerade problem:} 
modern EoS for cold, high density hadronic and quark matter are almost identical \cite{Alford:2004pf}; and 
\item \underline{the reconfinement puzzle:} 
what to do when after a deconfinement transition the hadronic EoS becomes favourable again
\cite{Lastowiecki:2011hh,Zdunik:2012dj}?
\end{itemize} 
In the left panel of Fig.~\ref{fig:1} we illustrate this situation by adapting a viewgraph from 
Ref.~\cite{Zdunik:2012dj}.

Within the present  contribution, we discuss that possibly the resolution of these puzzles can be given by accounting for the compositeness of baryons. 
In contrast to quarks and leptons, which are elementary and thus pointlike objects,  hadrons and in particular  baryons are bound states of quarks and have a finite size. In spite of this, within the quantum many-body theories of dense matter  they are usually treated  as  strongly interacting  pointlike particles. 
In order to account for the effect of the baryons' finite size  on the  EoS of  CS cores we  use a Lorentz invariant formulation of the  excluded volume approximation (EVA) \cite{Benic:2014jia}.
Depending on the phenomenological baryon size parameter the resulting stiffening of the hadronic EoS
may well remove all three puzzles mentioned above, see the right panel of Fig.~ \ref{fig:1}.
A sufficiently strong stiffening of the baryonic EoS can overcompensate the relative softening due to the occurrence of either hyperons or quark matter or both sequentially and still allow for fulfilling the new $2~$M$_\odot$ mass constraint on the one hand (no hyperon puzzle and no reconfinement).
On the other hand, if the quark matter EoS was already as stiff as that for pointlike baryons (masquerade) this additional stiffening due to the EVA will remove the masquerade effect and make the deconfinement transition strongly first order.

It is interesting to note that exactly such a type of EoS has been suggested recently \cite{Benic:2014jia}
when attempting
to answer the question: Can a strong first order phase transition in the CS interior be recognized from a
measurement of mass and radius for a sufficient (but most likely small) number of CS ?
The answer is positive and leads to the characteristic feature of high-mass twin stars, see also
\cite{Alvarez-Castillo:2013cxa,Blaschke:2013ana}.
In the following two sections we shall discuss the EVA and the density-dependent stiffening of quark matter in an advanced Nambu--Jona-Lasinio (NJL) model, respectively, both being the main ingredients to this new class of EoS for CS to be summarized in the subsequent section. After that, we outline the modern CS observations which shall be used to extract the most probable parametrization of the high-density EoS applying Bayesian analysis techniques \cite{Alvarez-Castillo:2014xea,Alvarez-Castillo:2014nua}. 
Finally, we present our conclusions. 

\section{Excluded volume approximation to quark level Pauli blocking in nuclear matter}

The finite-particle-size effect becomes  generally important   in many-particle systems as soon as the dimension of a particle is no longer much smaller than  an average interparticle distance. A classical example is the construction of a Van der Waals  model  for  non-ideal  gases (see, e.g., \cite{Landau:1999ft}).  
The finite size of ions is also considered  in the thermodynamics of dense  Coulomb plasma. It was included in the model of dense hot matter at subnuclear density when the EOS  of a collapsing supernova core was studied \cite{Lattimer:1985zf,Lattimer:1991nc} .  
The finite size of  hadrons is important  for the correct description of hot dense plasmas formed in heavy-ion  collisions, especially in the context of the transition to a quark gluon plasma 
\cite{Rischke:1991ke,Satarov:2009zx,Begun:2012rf}.  

It has been demonstrated before \cite{Ropke:1986qs} that a hard-core type repulsion between nucleons would result from the account of the Pauli principle for quarks from different hadrons in complete analogy to the molecular hard-sphere repulsion stems from the Pauli blocking between electron orbitals. 
Around nuclear saturation density, the resulting positive energy shift compares very well with the 
repulsive part of the effective Skyrme functional of \cite{Vautherin:1971aw}, see also \cite{Blaschke:1988zt}.
The resulting stiffening of the equation of state at high densities allowed to discuss high mass neutron and hybrid stars \cite{Blaschke:1989nn}. 
At present a reconsideration of the Pauli blocking between nucleons in dense nuclear matter is under way
\cite{Blaschke:2015} which takes into account the chiral restoration transition and partial delocalization of 
quark wave functions, thereby extending the context of a relativistic meanfield (RMF) theory of nuclear matter.
It is interesting to note that such a hybrid approach which implements the chirally improved Pauli blocking effect into a RMF approach to nuclear matter leads to an EoS for CS that is similar to a density dependent
RMF model (DD2 \cite{Typel:1999yq} with the parametrization given in \cite{Typel:2009sy}) generalized by the inclusion of the EVA, see \cite{Benic:2014jia}. 
Rooting the EVA in the Pauli principle as a consequence of symmetry principles in many-fermion systems with bound states such as clustered quark matter allows to understand the corresponding delocalization of the quark cluster wave functions in dense CS matter as a precursor effect signalling the onset of quark deconfinement. 
A modern description of quark matter itself under CS conditions, however, requires chiral quark model approaches such as the NJL type models capturing the essential effect of chiral symmetry restoration at high densities, eventually also the occurrence of colour superconducting phases, see 
\cite{Buballa:2003qv} for a review.

\section{Advanced NJL model for high-density quark matter}

Advanced NJL quark matter models have become the modern state-of-the art in describing deconfined quark matter in CS since they provide a microscopic and robust description of the phenomenon of dynamical chiral symmetry breaking in the quark sector of QCD. 
The description of normal quark matter phases with two and three flavours is straightforward and shows the qualitative difference to bag models of quark matter already: 
light and strange quark flavours appear sequentially \cite{Blaschke:2008br} due to the fact that different chemical potentials are required to reach the dynamically generated light and strange quark masses, so that strange quarks appear in cold quark matter only at high densities, eventually too high for CS interiors
\cite{Klahn:2006iw,Blaschke:2010vd,Klahn:2013kga}. 

Under neutron star constraints, however, it is customary to discuss color superconducting quark matter phases which appear when bosonic diquarks (Cooper pairs) form a  condensate in relevant interaction channels (Alford et al. 2008). 
The solution of the three-flavour colour superconducting (3FCS) NJL model requires advanced numerical techniques for the selfconsistent solution of the coupled set of gap equations for masses, diquark gaps. 
It has been pioneered in 2005 independently by two groups, Blaschke et al. \cite{Blaschke:2005uj} and Ruester et al. \cite{Ruester:2005jc} who showed that all previous results for superconducting quark matter in CS interiors (which worked with fixed strange quark mass) were flawed since they could not reveal the feature of sequential deconfinement of strangeness. 
This code has been developed further to include vector meanfields which provide a stiffening of high-density quark matter and allow to fulfil high-mass pulsar constraints \cite{Klahn:2006iw,Blaschke:2010vd,Klahn:2013kga}. 
It has been further developed to include the coupling to the Polyakov loop (PNJL model) as an implementation of confining effects at finite temperatures to facilitate applications to supernova collapse simulations \cite{Fischer:2011zj}. 
It is the "workhorse" for studies of quark matter in CS interiors, available to the Wroclaw group. 
In order to model high mass hybrid stars it has proven successful to extend the PNJL quark matter description by  higher order quark interactions equivalent to quartic selfinteractions of mesonic meanfields
\cite{Benic:2014iaa}.

\begin{figure}
  \includegraphics[height=.35\textheight]{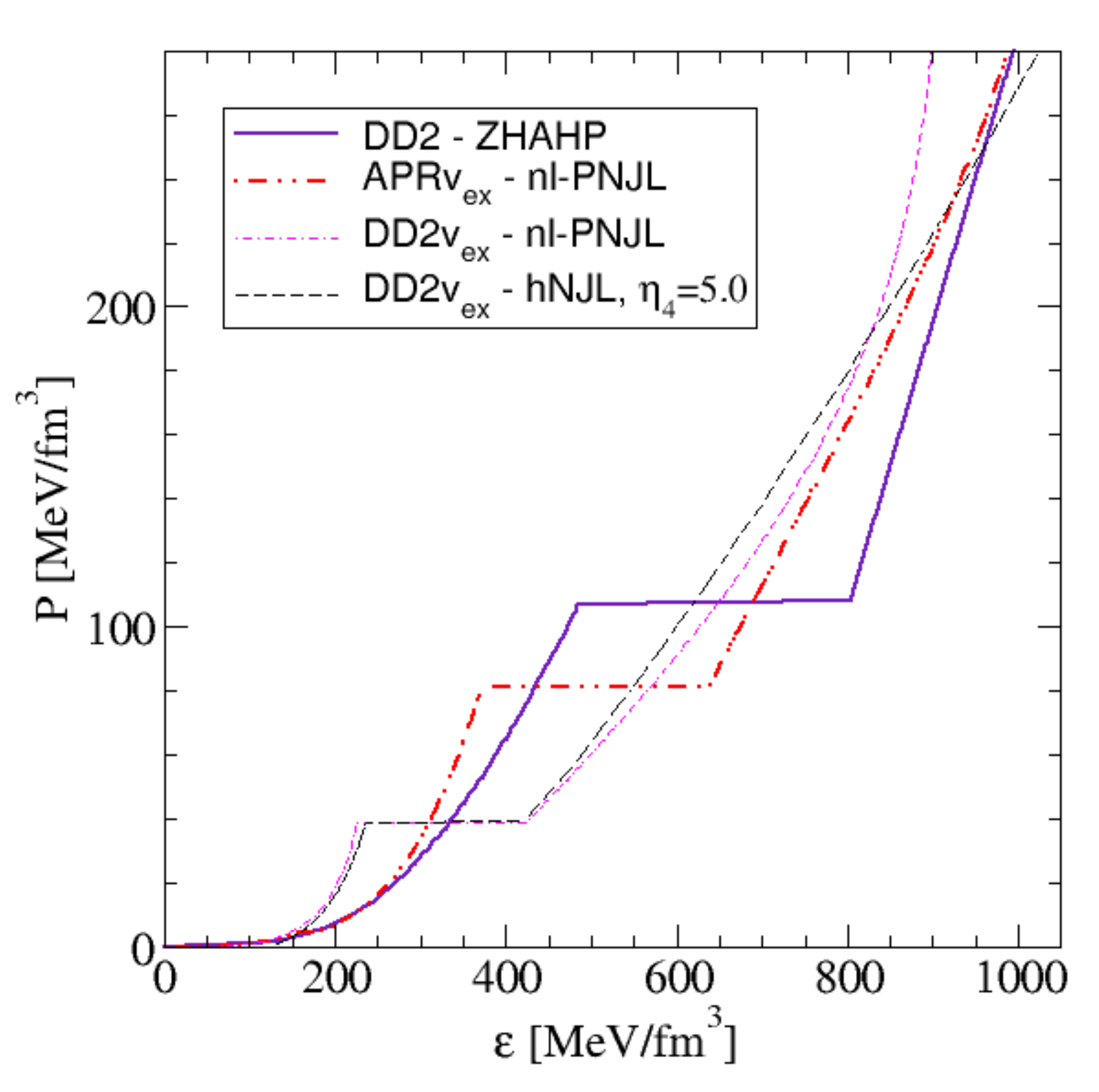}
  \includegraphics[height=.35\textheight]{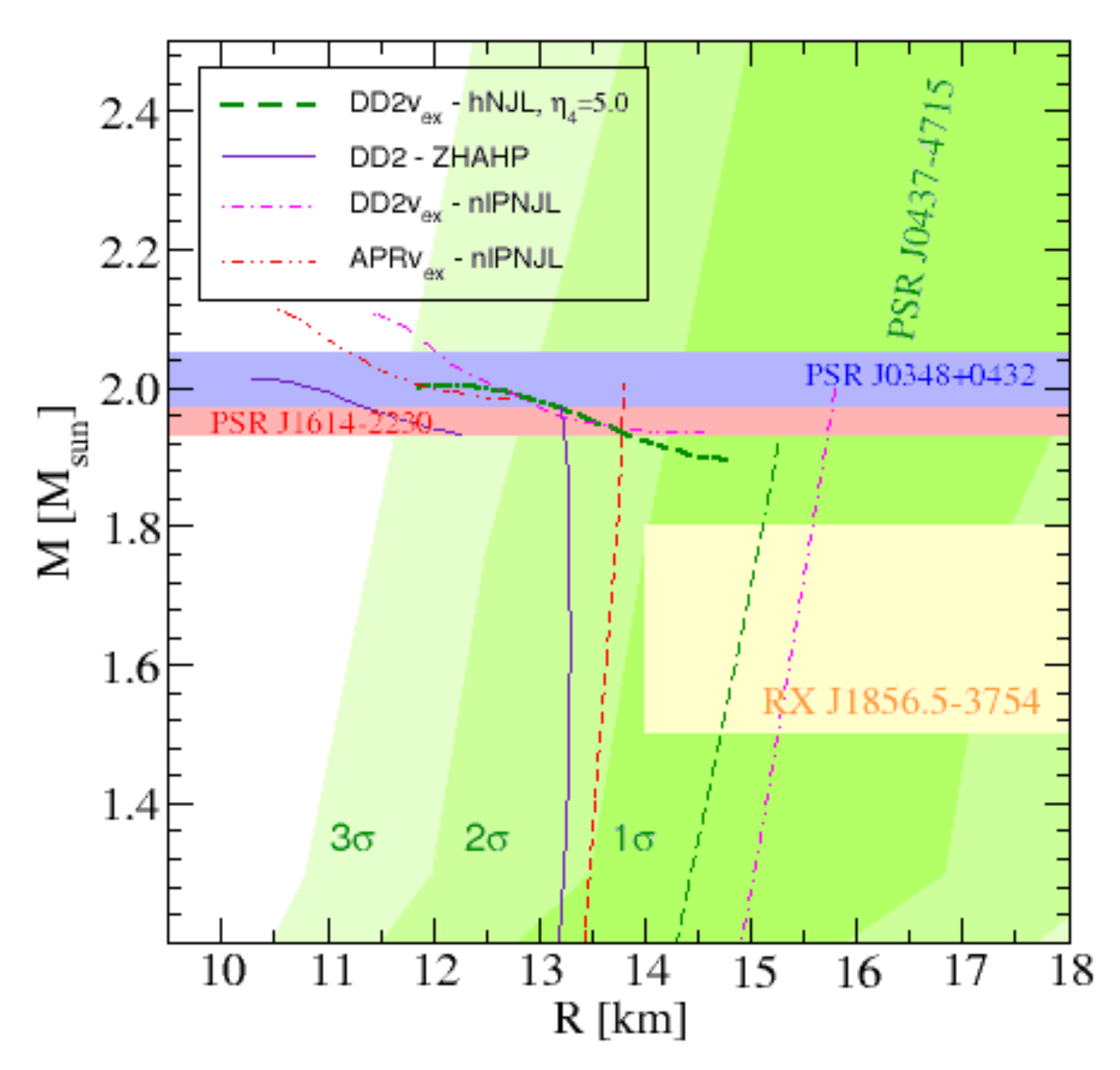}
  \caption{Left panel: Different realizations of EVA stiffening of the hadronic EoS with a Maxwell construction first order phase transition to stiff quark matter, for details see text. 
  Right panel: Solutions of the TOV equations for mass-radius relationships corresponding to the EoS shown in the left panel. All solutions have in common that they produce a "third family" of quark-hadron hybrid stars at high mass and that the "second family" of neutron stars has rather large radii as supported
  by recent pulsar timing analyses. 
  \label{fig:2}}
\end{figure}

In \cite{Benic:2014jia} a phase transition to the EVA improved DD2 EoS has been constructed and the corresponding compact star sequences in the $M-R$ diagram  have been obtained, see Fig.~\ref{fig:2}.
These sequences shown in the right panel of that figure elucidate the consequence of the density dependent stiffening of the quark matter EoS resulting from the higher order quark coupling in the vector channel: the onset of quark matter is shifted to higher star masses and eventually the conversion of the hadronic to a quark matter interior is followed by an unstable branch which is indicated by a break in the $M-R$ sequence. Due to the sufficient stiffening of high density quark matter, the stability is recovered for hybrid stars with quark matter interior, also at high mass of about $2~$M$_\odot$, but with significantly 
smaller radii. This third family of CS is observationally detectable (in principle) by simultaneous mass and radius measurement with sufficient accuracy whereby the high masses should be about the same, such as for the two known high-mass pulsars  PSR J1614-2230  \cite{Demorest:2010bx} and 
PSR J0348+0432 \cite{Antoniadis:2013pzd} but the radii should differ, resulting in the "mass twin" 
phenomenon \cite{Gerlach:1968zz,Kampfer:1981yr,Schertler:2000xq,Glendenning:1998ag}.

Also shown are previous models which could obtain the high mass twin phenomenon, see \cite{Blaschke:2013ana} for comparison.
In the left panel of Fig.~\ref{fig:2} we show the corresponding EoS that would exhibit the high mass twin phenomenon. They are all characterized by a strong first order phase transition where actually the jump
in energy density at the transition (so-called "latent heat") amounts to at least 60\% of the critical 
energy density at the transition.
Note also that the microscopically motivated high mass twin EoS which we emphasize in this work has a rather low critical pressure, below $100~{\rm MeV/fm}^3$. 
It is worth mentioning in this context that statistical model analyses of the chemical freeze-out in ultrarelativistic heavy-ion collisions find indications for a universal pressure along the freeze-out line with 
a value of about  $80~{\rm MeV/fm}^3$ \cite{Petran:2013qla}.
Another interesting aspect of this universality is that under completely different conditions in ab-initio studies of finite temperature QCD on the lattice the range of pressures in the crossover transition region has been found to be $40 - 80~{\rm MeV/fm}^3$ \cite{Bazavov:2014pvz}, see also Fig.~1 of 
Ref.~\cite{Alvarez-Castillo:2014dva}.
In \cite{Alvarez-Castillo:2014dva} it has been investigated how robust the high mass twin phenomenon would be against "smoothing" the phase transition as it occurs, e.g., when the formation of structures (so-called "pasta phases") in the phase coexistence region is considered. 
The result is that the observation of high mass twin stars would allow rather robustly the conclusion that there is a first order phase transition in the CS interior, likely with the formation of pasta. Such an observation would then allow the conclusion that in the QCD phase diagram there should exist at least one critical endpoint (CEP) of first order phase transitions 
\cite{Benic:2014jia,Alvarez-Castillo:2013cxa,Blaschke:2013ana}. 
This would be a very reassuring message for upcoming experimental programs for exploring the possibility of a  CEP and the existence of a mixed phase in heavy ion collisions, e.g., at 
FAIR CBM \cite{Friman:2011zz} and 
NICA MPD \cite{nica:2015x}.
In order to answer the question how precise mass and radius measurements must be for detecting unambiguously the existence of high mass twins in the $M-R$ diagram a new Bayesian analysis scheme has been suggested \cite{Alvarez-Castillo:2014xea,Alvarez-Castillo:2014nua} which is at considerable
difference to the previously suggested scheme \cite{Steiner:2010fz,Steiner:2012xt}.

\section{New Bayesian analysis}

The Bayesian analysis method is a well developed tool in physical research when parametrically known laws of nature shall be tested against experimental data. 
Recently this tool has been applied to the question of the inversion of the Tolman-Oppenheimer-Volkoff (TOV) equation for CS structure and stability, i.e. to obtain a most probable equation of state for NS matter from observational data (with error bands) for CS in the mass-radius diagram \cite{Steiner:2010fz,Steiner:2012xt}. 
This first application of Bayesian analysis to the TOV inversion problem is, however, problematic as it uses burst sources and those for which we have only information about the X-ray spectrum, not in other wavelength bands such as the optical. 
The lesson from the nearby X-ray dim lonely neutron star RXJ 1856.5-3754 is, however, that without the optical part of the spectrum a "hot spot" on the surface may be mistaken for the whole emitting surface so that extracted radii a grossly underestimated \cite{Trumper:2003we,Trumper:2011}!     
Therefore we have suggested a new set of observations to be used for the Bayesian analysis and explored this in a proof-of-principle application \cite{Alvarez-Castillo:2014xea,Alvarez-Castillo:2014nua} to the case of a schematic hybrid EoS of the type suggested by Zdunik \& Haensel \cite{Zdunik:2012dj} and by
Alford et al. \cite{Alford:2013aca}.  More work in this direction is in progress, actually employing the new class of CS EoS \cite{Benic:2014jia} which has two physical parameters to vary, the excluded volume for baryons stemming from the Pauli blocking on the quark level and the unknown coupling strength in the higher order quark coupling in the vector channel as discussed in this contribution.

\section{Conclusions}

In this contribution we have discussed that the resolution of three of the fundamental puzzles of CS structure, the hyperon puzzle, the masquerade problem and the reconfinement problem may likely be all solved by accounting for the compositeness of baryons (by excluded volume and/or quark Pauli blocking) on the hadronic side and by introducing stiffening effects on the quark matter side of the EoS.
The resulting new class of EoS for hybrid CS contains the interesting case of a strong first order phase transition which results in the high mass twin star phenomenon in the $M-R$ diagram, allowing for an astrophysical observation of the critical endpoint in the QCD phase diagram, long sought-for in ultrarelativistic heavy-ion collisions and lattice QCD simulations.

%%%%%%%%%%%%%%%%%%%%%%%%%%%%%%%%%%%%%%%%%%%%%%%%
%% BACKMATTER
%%%%%%%%%%%%%%%%%%%%%%%%%%%%%%%%%%%%%%%%%%%%%%%%

\begin{theacknowledgments}
D.B. is grateful to T. Fischer, H. Grigorian, P. Haensel, T. Kl\"ahn and L. Zdunik for many interesting discussions of the puzzles and the suggested solution considered in this contribution.
D.E.A-C. acknowledges support by the programmes for exchange between JINR Dubna and German Institutes (Heisenberg-Landau programme) as well as Polish Institutes (Bogoliubov-Infeld programme). This work was supported in part by the Polish National Science Center (NCN) under grant number 
2014/13/B/ST9/02621.
%UMO-2011/02/A/ST2/00306. 
The authors gratefully acknowledge the COST Action MP1304 "NewCompStar" for supporting their networking and collaboration activities.
\end{theacknowledgments}
%%%%%%%%%%%%%%%%%%%%%%%%%%%%%%%%%%%%%%%%%%%
%% The following lines show an example how to produce a bibliography
%% without the help of the BibTeX program. This could be used instead
%% of the above.
%%%%%%%%%%%%%%%%%%%%%%%%%%%%%%%%%%%%%%%%%%%

\end{document}